\documentclass[letterpaper, 12pt]{article} 
\usepackage[T1]{fontenc}
\usepackage[latin1]{inputenc}
\usepackage{amsmath}
\usepackage{amssymb}
\usepackage{tabularx}
\usepackage{graphicx}
\usepackage{lscape}
\usepackage{longtable}
\usepackage[final]{pdfpages}
\usepackage{afterpage}
\usepackage{setspace}
\usepackage{float}
\usepackage{comment}
\usepackage{multicol}
\usepackage{multirow}
\usepackage{booktabs}
\usepackage{natbib}

\date{}

\usepackage{subcaption}
\usepackage{authblk}

\title{Truck Traffic Monitoring with Satellite Images}

\author[1, 2]{Lynn H.~Kaack}
\author[3]{George H.~Chen}
\author[1]{M.~Granger Morgan}
\small{
\affil[1]{\small{Department of Engineering and Public Policy, Carnegie Mellon University}}
\affil[2]{Energy Politics Group, ETH Z\"urich}
\affil[3]{Heinz College of Information Systems and Public Policy, Carnegie Mellon University}}

\begin{document}

\maketitle

\begin{abstract}
The road freight sector is responsible for a large and growing share of greenhouse gas emissions, but reliable data on the amount of freight that is moved on roads in many parts of the world are scarce.  Many low- and middle-income countries have limited ground-based traffic monitoring and freight surveying activities. 
In this proof of concept, we show that we can use an object detection network to count trucks in satellite images and predict average annual daily truck traffic from those counts. 
 We describe a complete model, test the uncertainty of the estimation, and discuss the transfer to developing countries.
\end{abstract}

\section{Introduction}
Especially across the developing world, a key barrier to identifying opportunities for mitigating climate change is the lack of sufficiently granular, high-quality data. Heavy- and medium-duty trucking accounts for 7\% of total world energy-related CO$_2$ emissions \citep{iea2017future}, with much of the growth occurring in developing countries \citep{kaack2018decarbonizing}. In order to successfully implement policies and make targeted investments, reliable data about the volume of freight that is moved on roads is crucial. More than half of all countries do not collect national road freight activity data and where estimates exist, they are typically survey-based and often inadequate \citep{kaack2018decarbonizing}. 
Knowing truck movements is also important for a variety of economic analyses and for road maintenance planning, even if only based on short-duration counts \citep{fhwa2016mon}, but such ground-based traffic monitoring is costly and not performed in many countries. 

In this paper, we propose a remote sensing approach to 
monitor freight vehicles through the use of high-resolution satellite images. As satellite images become both cheaper and are taken at a higher resolution over time, we anticipate that our approach will become scalable at an affordable cost within the next few years to much larger geographic regions. We take advantage of recent advances in convolutional neural networks for object detection. These methods have already been successfully applied to detecting vehicles in satellite images \citep{sommer2017fast, jiang2015deep, chen2014vehicle, deng2017toward, mundhenk2016large}. Most work has focused on cars, and to a lesser extent on multiple vehicle classes including trucks \citep{liu2015fast, sommer2017fast}. Note that a satellite image is only for a single snapshot in time, whereas conventional traffic estimates are taken 
over a much longer period of time. Thus, in our approach we must separately model how traffic changes with time.

We begin by providing a brief overview of traditional ground-based traffic monitoring and remote sensing alternatives (Section \ref{sec:background}). We then introduce our framework, which consists of a truck detection model and a temporal traffic monitoring model (Section~\ref{sec:framework}). For training the detection model, we have curated our own dataset. We validate and test our approach using data from the New York Thruway and California and assess how the model transfers to data from Brazil (Section~\ref{sec:exp}). We conclude with a qualitative discussion of the results and future work (Section~\ref{sec:discussion}).

\section{Background: Traffic monitoring and freight surveying}
\label{sec:background}

The US Federal Highway Administration (FHWA) highlights the importance of vehicle counting for traffic monitoring, as counting provides statistics such as the Annual Average Daily Truck Traffic (AADTT) \citep{fhwa2016mon}.
Ground-based automatic vehicle counting devices include pneumatic tubes, inductive loop detectors, magnetic sensors, 
video detection systems, and several others. Installation and maintenance for some of these systems requires pavement cuts and lane closures.
Traffic monitoring is usually based on continuous counts, which also provide the basis for periodic (e.g., hour of the day) factors applied to short duration counts. Typical short duration detection periods are between 24 hours and a week long \citep{fhwa2016mon}.

\paragraph{Traffic monitoring with remote sensing.}
As ground-based detection devices can be prone to failure and are too costly to install and maintain in some countries, there is a need for alternative monitoring technologies, such as through GPS data from cell phones \citep{herrera2010evaluation} or with aerial or high-resolution satellite images, and even lower-resolution satellite images \citep{larsen2009traffic, eikvil2009classification}. There is also potential for using drones \citep{kanistras2015survey}.
With remote sensing, a large number of roads can be covered at the same instance, many of which are not equipped with costly sensors \citep{larsen2009traffic, eikvil2009classification} (e.g., rural or remote roads).
Also, areas that are difficult to access, for example due to a disaster or conflict, could be monitored \citep{gerhardinger2005vehicles}.
A weakness of the method is that traffic fluctuations on short time scales as well as time-of-day, day-of-week, and seasonal traffic patterns can distort the accuracy of the estimate of the AADTT \citep{fhwa2016mon}, and such images are only available for daylight and under cloud-free conditions. 
In addition, this method requires advanced analytical and computational resources.
The uptake of remote sensing methods for transportation applications has been slow but it promises to offer cost-effective and scalable options for a multitude of applications \citep{bridgelall2016remote, bowen2004assessment}.

\paragraph{Freight surveying.}
Data on road freight activity, measured in tonne-km, are typically obtained through costly national surveys of shipping companies, which need to provide information on origin, destination, weight, and other indicators of all shipments. 
Less than half of the countries in the world collect this type of information \citep{kaack2018decarbonizing}. Truck counts can be used to estimate the freight activity \citep{fhwa2016mon}.

\section{Framework}\label{sec:framework}

\begin{figure}[h]
\centering
   \includegraphics[scale=0.55]{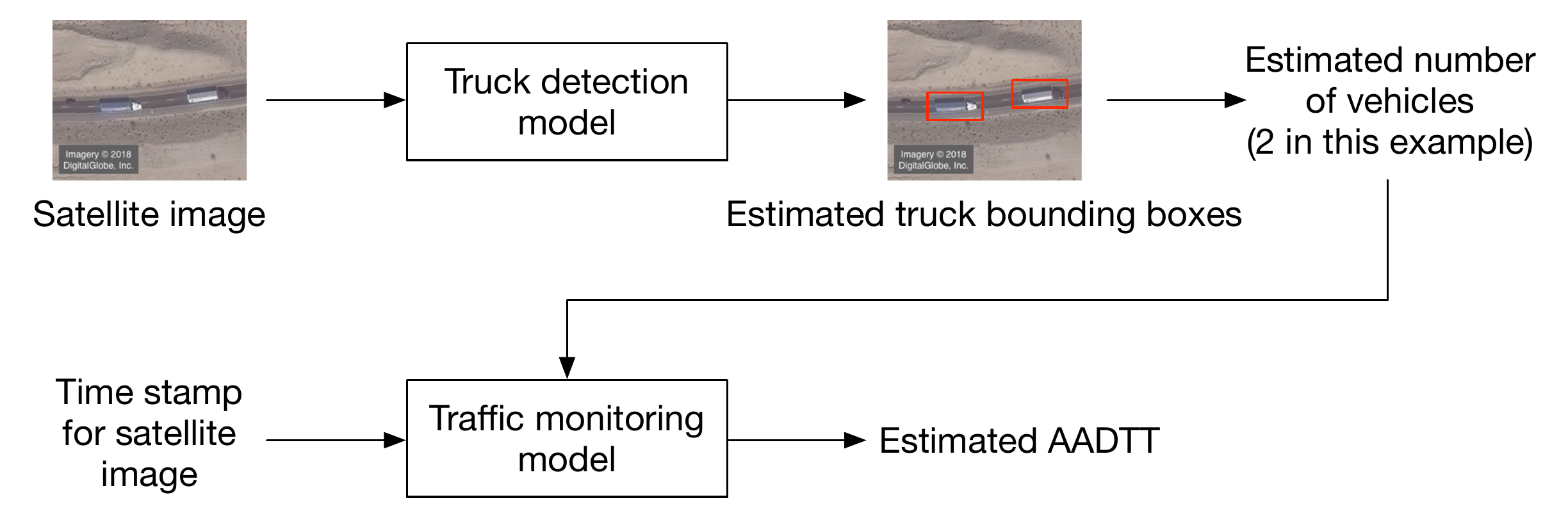} 
\caption{A diagram illustrating the overall framework we use for predicting the AADTT from satellite images.}
\label{fig:diagr1}
\end{figure}

Our framework consists of a truck detection model and traffic monitoring model (Fig.~\ref{fig:diagr1}). The detection model counts the number of freight vehicles on roads in a satellite image, and the monitoring model translates these counts into an AADTT estimate. 
The traffic  monitoring model takes as input an estimated vehicle count (obtained from the satellite image via the truck detection model) along with the timestamp for the satellite image; the traffic  monitoring model's output is an AADTT estimate, which is a vehicle count averaged over time rather than for a single snapshot in time.

\subsection{Truck detection model}
\label{sec:detection-model}

For detecting trucks, we combine a convolutional neural network that does object detection with existing geospatial data to filter out trucks that are not near roads.

\paragraph{Object detector.}
The object detection model provides the vehicle count from an image. 
\cite{huang2017speed} identified three object detection meta-architectures, which are Faster Region-based Convolutional Neural Networks (Faster R-CNN), Single Shot Detectors (SSD) and Region-based Fully Convolutional Networks (R-FCN). They have tested models based on these meta-architectures for speed and accuracy, and have found that Faster R-CNN often achieved the highest accuracy, while SSD excelled in speed.
Faster R-CNN first proposes regions with the Region Proposal Network (RPN) and then uses the Fast R-CNN detector \citep{girshick2015fast} for object detection, sharing convolutional layers. 
While Faster R-CNN first classifies the objectiveness of proposed boxes, and then predicts the class in another network, SSD directly classifies and regresses boxes, which makes it much faster to train and perform inference. 
We compare Faster R-CNN \citep{ren2015faster} with 50- and 101-layer Resnets \citep{he2016deep} and SSD Inception V2 \citep{liu2016ssd} for our application.
We use the default implementations for the COCO image dataset from the Tensorflow Object Detection API \citep{huang2017speed} 
and pre-trained convolutional layers. We count a truck as detected if its bounding box has an intersection over union (IoU) with the ground truth box of at least 0.3.

\paragraph{Road filter.}
We only want to count trucks that are driving a specific road of interest, and exclude those that are sitting in parking lots or traveling on other nearby roads. To filter out irrelevant predictions from the detection model, we use geospatial data. Those data are ubiquitous, and also available for main transit highways in developing countries. We count a truck if at least one corner of its bounding box is within a certain distance of the center of the road. If both lanes are indicated, we set this distance to 8 meters, which approximately accommodates a four-lane highway. 
This filter is applied to both the ground truth validation and test datasets and the predictions from the model.

\subsection{Traffic  monitoring model}
\label{sec:monitoring-model}

To use a snapshot image to approximate ground-based annual average daily values, we assume that all $c_I$ vehicles travel with a constant speed $v$ within the interval $s$ defined by the geographic extent of the image or the length of the highway section. From that we infer the time $t_I(v, s)$ that it takes for a vehicle to travel from the start to the end point in the interval. A detector installed in the end point should count $c_I$ vehicles in time $t_I(v, s)$. 
The FHWA recommends that traffic density variation factors\footnote{The traffic density variation factor is specific to the hour of the day $h\in \{1,2, ..., 24\}$, the day of the week $d \in \{1, 2, ..., 7\}$, and the month of the year $m \in \{1, 2, ..., 12\}$.} $f_{h,d,m}$ be applied when using less-than-a-day counts to compute the AADTT \citep{fhwa2016mon}, so as to account for time-of-day, day-of-the-week (DOW) and monthly variations. 
We can approximate the average daily (bidirectional) counts as
\begin{equation}
AADTT \approx \frac{c_I}{t_I(v, s)} \cdot \frac1{f_{h,d,m}}, 
\label{eq:AADTT}
\end{equation}
which is reported in units of \emph{number of vehicles per day} (the factor $f_{h,d,m}$ is dimensionless).
Detailed information about traffic patterns 
can reduce the error of the estimate. Here, we assume that no information about traffic variation in the test region is given, and we need to approximate the factors and their uncertainty from regions where truck traffic is monitored. 

The factor $f_{h,d,m}$ is estimated as the conditional average of normalized hourly count values. We created normalized count values by dividing the hourly count data by the mean of all hourly counts in the year. 
We compared random forest regression and six 
different linear models using factors that can capture seasonal effects:
\begin{enumerate}
\item normalized count $\sim$ weekend + hour (factor)
\item normalized count $\sim$ DOW (factor) + daytime
\item normalized count $\sim$ DOW + hour
\item normalized count $\sim$ DOW + hour + hour * DOW
\item normalized count $\sim$ month (factor) + DOW + hour
\item normalized count $\sim$ month + DOW + hour + \\
hour * DOW.
\end{enumerate}
These models are informed by the recommended practices of the FHWA \citep{fhwa2016mon, krile2016assessing}.

\paragraph{Uncertainty analysis.}
We use a Monte Carlo method to estimate the uncertainty of the traffic monitoring model based on the uncertainty of speed and time variation factors. The basic idea is that for a specific satellite image/highway section, we randomly sample speed $v$ and time variation factors $f_{h,d,m}$, and then compute the AADTT estimate as given by Equation \eqref{eq:AADTT}. Repeating this procedure many times, we obtain a distribution for the model's AADTT estimate for the specific road section of interest. We use the median and interquartile range of this distribution as robust final estimates of the AADTT and its uncertainty.

We generate the random speed and time variation factors as follows. For speed $v$, we use a heuristic and sample from a Gaussian with mean $v_0$ and standard deviation $0.05 v_0$ (see Sections \ref{sec:dataUS} and \ref{sec:dataBR} for assumptions on $v_0$).
For the time-variation factors $f_{h,d,m}$, we use a non-parametric approach: we sample from the subset of training residuals from estimating $f_{h,d,m}$ that have the same DOW and hour as the test time stamp. Constraining the sample of residuals in that way approximately ensures that the factor uncertainty is not negative. Future work may focus on constructing statistically valid prediction intervals (for example using conformal inference \citep{doi:10.1080/01621459.2017.1307116}).
By making assumptions about the distribution of payloads of the freight vehicles \citep{fhwa2016mon}, one could use this approach to further estimate the freight activity through truck counts.

\subsection{Training procedure}
\begin{figure}[t]
\centering
   \includegraphics[scale=0.55]{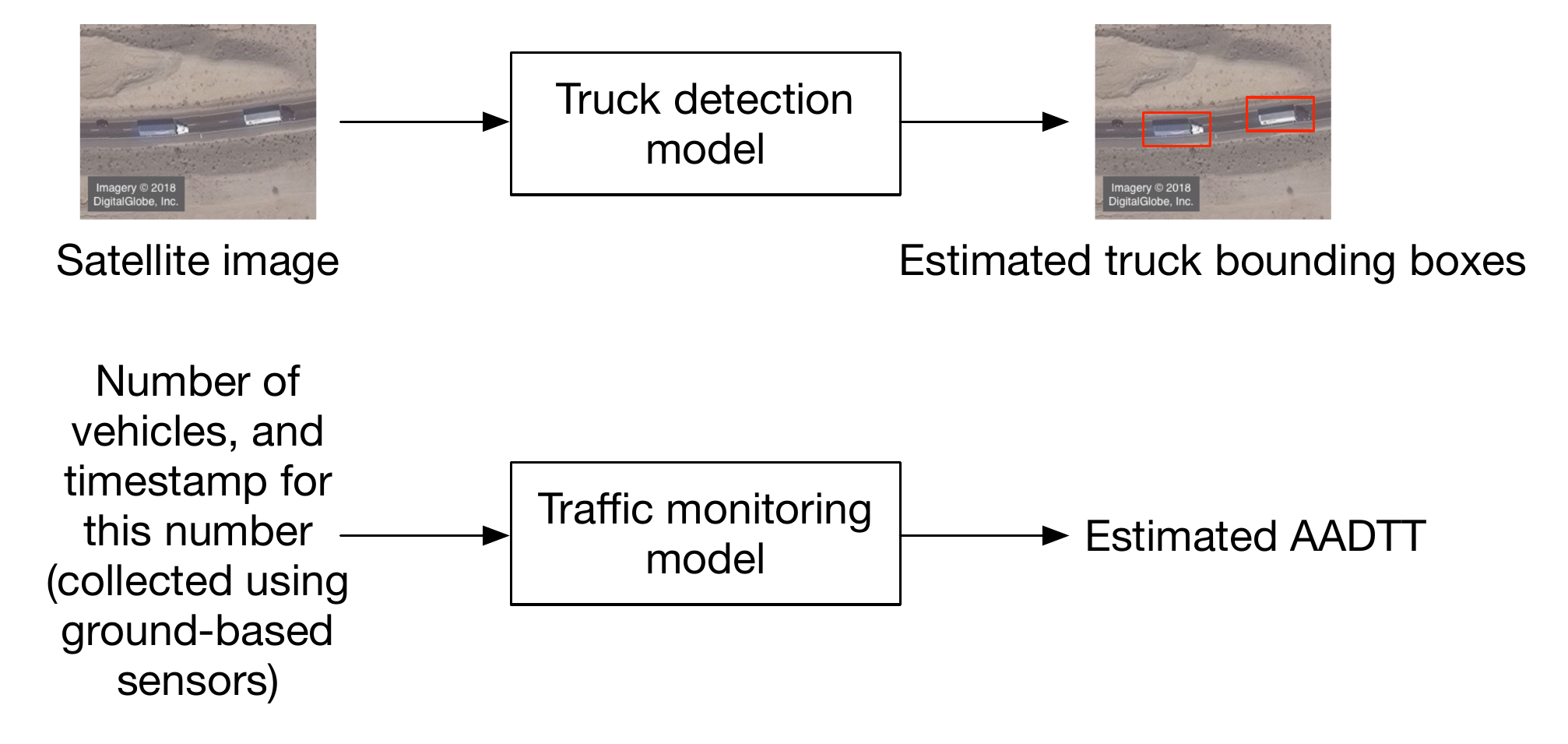} 
\caption{An illustration of the training procedure. We train and select parameters for the two submodels separately.}
\label{fig:diagr2}
\end{figure}

We train and select the best parameters for the truck detection and traffic monitoring components separately (Fig.~\ref{fig:diagr2}). For both components, we use a validation dataset for model and parameter selection.

The truck detection model is trained using annotated satellite images with known truck bounding boxes (and hence known truck counts). As mentioned previously, we train three different object detectors: Faster R-CNN Resnet 50, Faster R-CNN Resnet 101, and SSD Inception V2. All three detectors have a prediction probability threshold parameter that trades off between true and false positive rates. We select this threshold using a validation set by minimizing a prediction error of total truck counts per image (which can include true positives and false positives). Specifically, the error we use is the average absolute count error over all validation images as the weighted sum of the relative absolute count error of each of $N$ images:
\begin{align}\label{eq:main}
\epsilon_{count} 
&= \frac{\sum_{i=1}^N |c_{pred}^{(i)} -c_{true}^{(i)}|}{\sum_{i=1}^N c_{true}^{(i)}},
\end{align}
where $c^{(i)}$ is the number of trucks in image $i$. While the unweighted mean absolute count error is often used \citep{Marsden:2017aa}, we chose to use a weighted sum to account for the relative importance of images with a lot of truck examples.

The traffic  monitoring model is trained using vehicle counts from ground-based sensors for the same spatial region covered by a satellite image.  
The satellite image with its estimated vehicle count from the truck detection model would only provide the vehicle count for a single time stamp, whereas ground-based sensors provide hourly vehicle counts. Having vehicle counts over time is crucial for training the traffic monitoring model because the model's output is an AADTT estimate, which measures traffic averaged over a whole year.

\section{Experiments}\label{sec:exp}
In a first experiment, we tested how well a model trained and validated on data from the NY Thruway could estimate the ground-based AADTT value on held-out sections of the Thruway and California. We then transferred the model to another country, where we tested the US-trained model on a highway in Brazil, and compared it to a model trained on local data.

\subsection{NY Thruway/CA}

\subsubsection{Data}\label{sec:dataUS}
We curated our own collection of 31cm-resolution, RGB-color satellite images provided by DigitalGlobe, Inc., (Appendix \ref{sec:ann}) since a large satellite image database ("xView" \citep{lam2018xview}) with several thousand labeled truck instances proved too inaccurate and other satellite image datasets contained only small numbers of trucks \citep{mundhenk2016large, razakarivony2016vehicle}. 
For training, we used images of several regions in the Northeastern US, primarily the NY Thruway, with a total of 2050 truck examples.\footnote{A smaller training set of approximately half the size showed similar performance.} For validation and parameter selection, we worked with another set of images from 4 sections of the NY Thruway, some partially covered by fog, that contain 340 truck examples (88 on road). 
For the road filter, we used shapefiles provided by the States of New York and California \citep{shape_thruway, shape_CA}.

For choosing the traffic monitoring regression models and to train the factors, we used hourly ground-based counts for four regions, namely the NY Thruway \citep{ezpass2017hourly, ezpass2016hourly}, California \citep{pemsTruck}, Brazil \citep{br_data}, and Germany \citep{bast2018Germany}. While the first are toll data, the latter three are datasets from short-term and continuous counters (refer to Appendix \ref{sec:mon}). 
We assumed a mean speed of 65 mi/hr for the Thruway and 70 mi/hr for California \citep{CA_speed}.

To assess the prediction accuracy of our overall framework, we used 11 test cases from the NY Thruway and 3 test cases from California, each consisting of a single image of unseen (in training/validation data) highway sections with a different time stamp. These test images showed a total of 541 trucks on the road.

\subsubsection{Truck detection model}

\begin{figure}[h]
   \centering
   \includegraphics[scale=0.6]{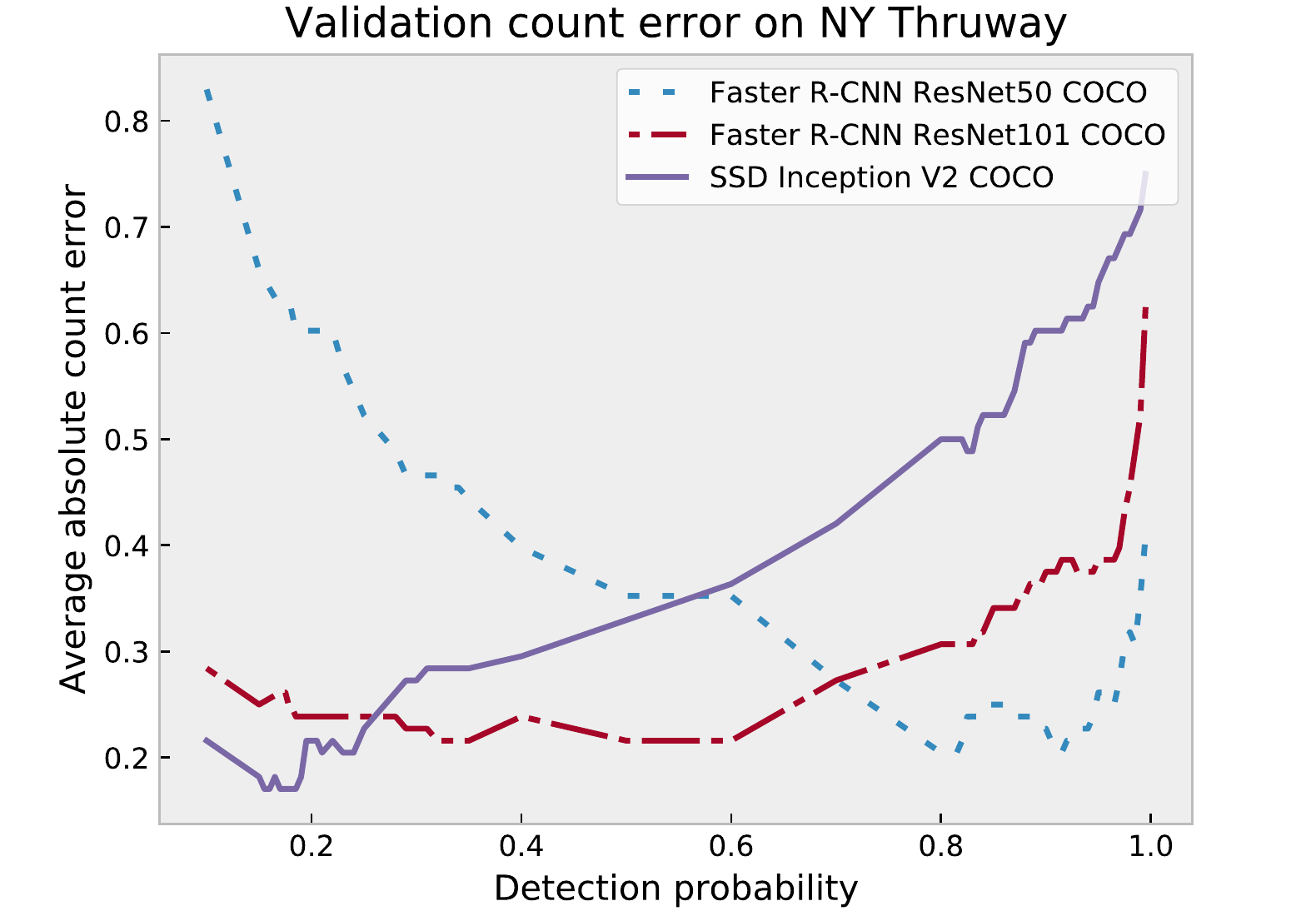}
     \caption{Error of truck counts, which also includes false positives, over a tuned probability parameter (detection prob.). For a high detection probability we only count very few trucks, which means that the model under-predicts and the error spikes. We see that the SSD achieved lowest count error.
     }
 \label{fig:count_opt}
 \end{figure}

 Fig.~\ref{fig:count_opt} shows the count error for various detection probability thresholds over the validation dataset (see Appendix \ref{sec:detailDet} for more validation results). We selected the model and the respective fixed detection probability threshold that minimized the count error over the validation images.  
 SSD Inception V2 achieved the lowest minimal count error on road, and higher precision, recall (Appendix \ref{sec:detailDet}) and speed. We chose to use SSD Inception V2 with prediction probability $p_{pred}=0.155$ to test the model.

\subsubsection{Traffic monitoring model}

We used regression to predict the time-varying factors for a test region where we do not have ground truth data. Using the uncertainty of these factors, a distribution of vehicle speeds, and the length of the section, we predicted the AADTT with the Monte Carlo method.

\paragraph{Factor model selection.}
Where the model is most useful, local vehicle counts data are not available, and hence we needed to find a model that predicts well using traffic-variation factors from other regions.
We used a cross validation procedure with models trained on ground-based counts from three regions and validated on a held-out fourth region.

For the cross validation, we selected the equivalent of 10 continuous ground-based counting stations from each region, where we prioritized those stations that have more data and a higher AADTT. Since some datasets contain short-term counts, to maintain approximate balance, we sampled more counting stations until we had as much data as 10 continuous counters or the dataset was exhausted (Appendix \ref{sec:countsumm}). Since the toll data for the NY Thruway are constrained to one single highway, we used only 6 toll booths here. We ignored inter-year variation. 

\begin{figure}[h]
\centering
   \includegraphics[scale=0.6]{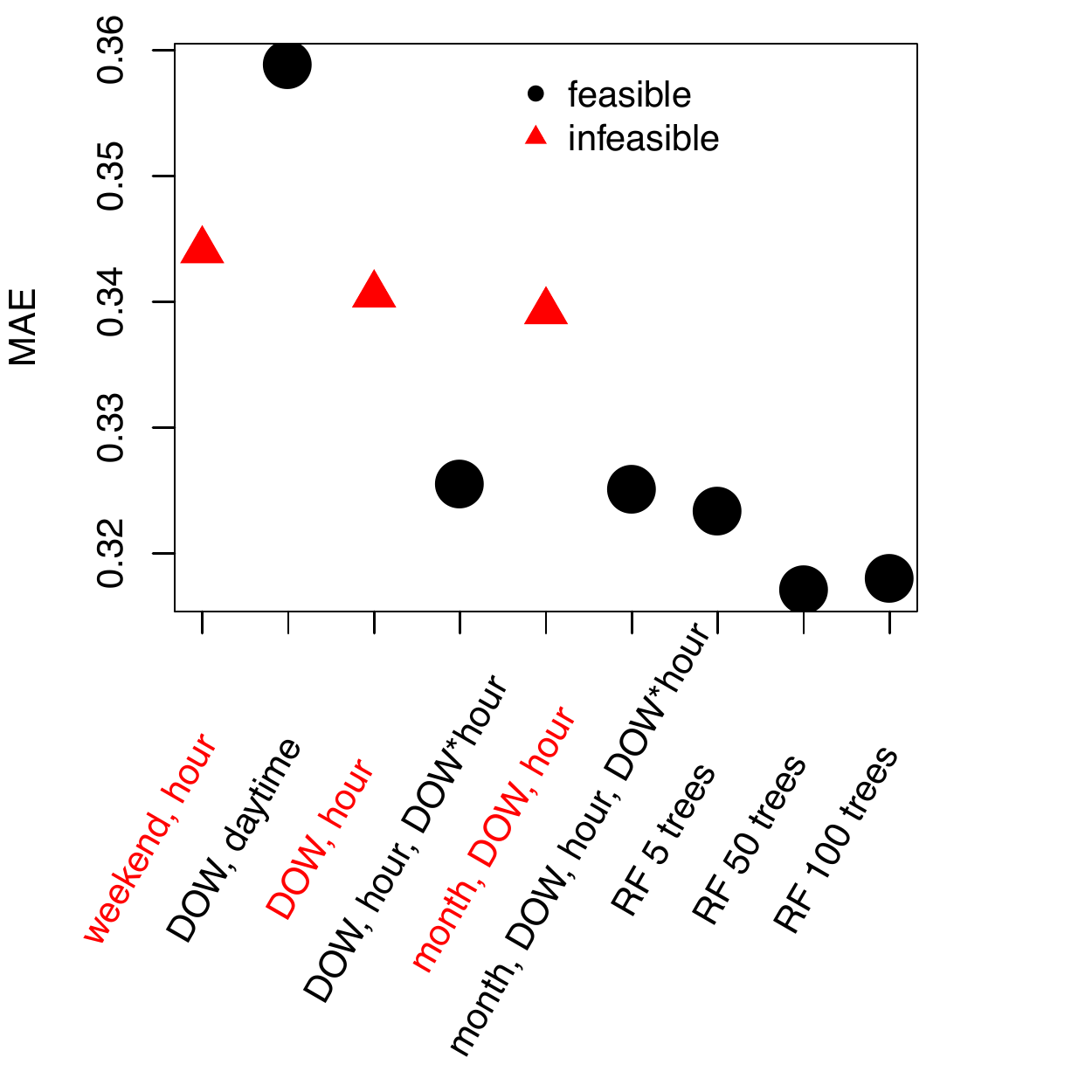}
\caption{Cross-validated MAE of different factor regression models to estimate the normalized hourly count, indicated by the independent variables. Some models predict infeasible (negative) values.
}
\label{fig:mse}
\end{figure}

We trained the model on three of the regions, using quantile regression for linear models for a robust estimate of the median, and recorded the mean absolute prediction error (MAE) 
on the held-out fourth region. The MAE 
averaged over all regions is reported in Fig.~\ref{fig:mse}. Some of the models, in particular those that do not have interactions that allow hourly patterns to differ between weekdays and weekend days, produced negative predictions. Since traffic counts are strictly positive, such predictions are infeasible, and we excluded these models. 
We found that the random forests outperformed even the most complex linear model, and chose a random forest with 50 trees to predict time-varying factors. For testing the whole model, the random forest was also trained on three other regions to test on a fourth.

\begin{figure}[h]
\centering
    \includegraphics[width=\linewidth]{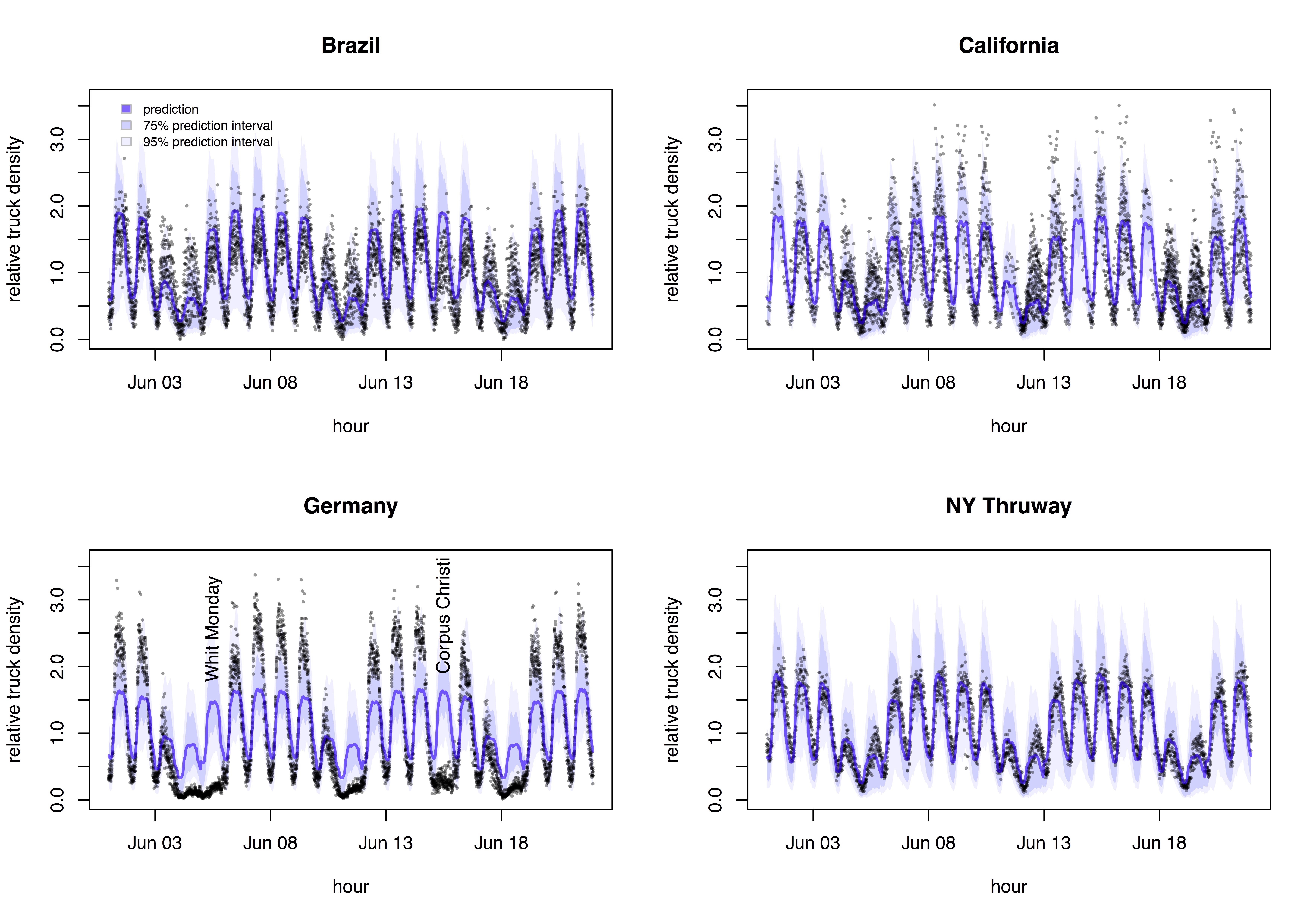} 
\caption{Out-of-sample traffic variability prediction for three example weeks in the four regions. Each plot shows the true normalized hourly truck count for all selected count stations as scattered points, and the prediction as a blue line, with non-parametric prediction intervals based on the training residuals. Each model was trained on hourly counts from all three other regions. 
Vertical text indicates German public holidays. Corpus Christi is noisy as it is not observed in all German states.}	
\label{fig:cvpred}
\end{figure}

Fig.~\ref{fig:cvpred} is a visualization of how the random forest regression predicts on each held-out region. Overall, the hourly pattern of truck traffic seems to transfer well to other regions. This example, however, also illuminates the shortcomings of the model. For example, Germany does not allow truck drivers to work on Sundays. We see that the particularly low values on Sundays in Germany, and the corresponding larger variability, were not well predicted by the fit on the other regions that do not have such strict labor rules. In return, the model trained with German data underpredicted Sundays in other regions. To improve the performance of the monitoring model, information about local labor rules may need to be incorporated. This figure also illustrates how such traffic models fail to predict deviations from regular patterns, such as holidays \citep{fhwa2016mon}.

\subsubsection{Test results and discussion}
\begin{figure}
\centering
   \includegraphics[width=0.95\linewidth]{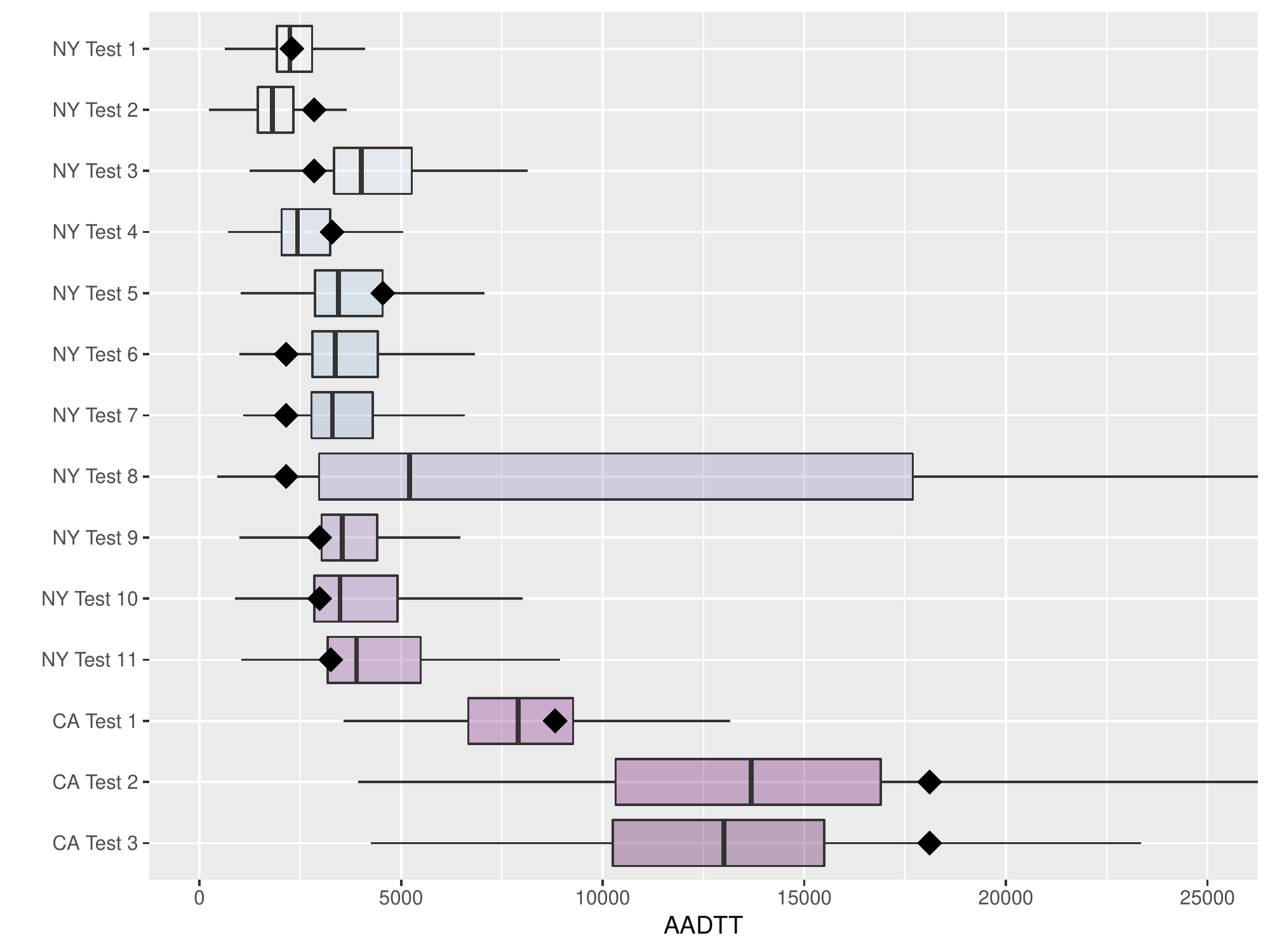} 
\caption{Predicted AADTT from satellite images (box plots) and ground-based AADTT (diamond) for different test regions on the NY Thruway and in California. 
NY Test 8 is based on an image taken on a Sunday.}
\label{fig:testresults}
\end{figure}

\begin{figure}[h]
\centering
   \includegraphics[scale=0.6]{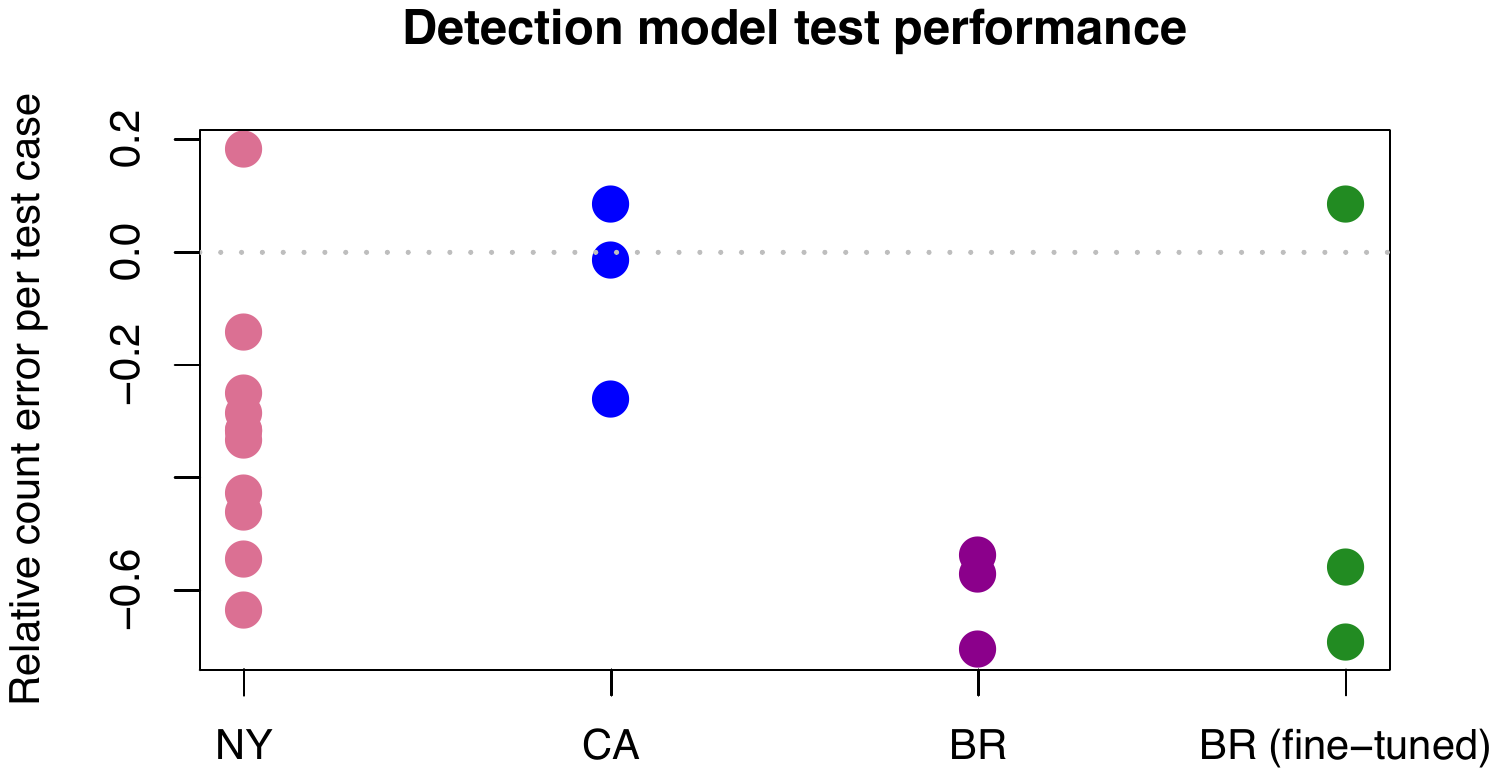}
\caption{Relative count error for each test case grouped by test regions. The first three regions correspond to the detection model trained on the Northeastern US. We see that the fine-tuned detection model improves detection performance for Brazil.}
\label{fig:detecTest}
\end{figure}

We used the predicted vehicle count from the detection model, the time stamp of the images, time-varying factors, and speed to make a probabilistic prediction of the AADTT. 
Fig.~\ref{fig:testresults} 
shows the AADTT estimates from our model compared to the values of traditional ground-based measurements. We found that in about half of the test cases, the ground-based AADTT was in or very close to the interquartile range of the predictions. This proof of concept does not include a statistical claim, as the sample size is small. 
The majority of truck detection model counts were lower than the true counts in the images (Fig.~\ref{fig:detecTest}). 
This suggests that improvements could be made by increasing the count accuracy of the detection model. Interestingly, the detection model trained on the Northeastern US performed well on California images, despite the different geography.  

Discrepancy in model performance is expected, given that a snapshot image corresponds to a single, very short counting time, and is sensitive to traffic fluctuations. We also find that the prediction for Sundays might be less precise (Test Case 8 is the only Sunday), as those days vary more between the regions and the traffic monitoring model is less accurate.

\subsection{Brazil}\label{sec:BR}
As our model is intended to make predictions in developing countries and emerging economies, we tested how well the model would perform if applied to Brazil. 
Brazil is suitable as it is an emerging economy but there are sufficient data available to analyze how well the model and each of its components generalize. 
We were particularly interested to test if additional fine-tuning of the detection model with local images would be necessary. 

\subsubsection{Data}\label{sec:dataBR}
For the detection model, we used images from DigitalGlobe, Inc. We annotated 2027 trucks for training, and curated separate validation data with images of two sections containing 409 trucks (119 on road).
For the monitoring model, we worked with traffic data from continuous and short term counters available through the Brazilian agency Departamento Nacional de Infraestrutura de Transportes (DNIT) \citep{br_data}. 
We tested on a section of the highway BR-116 between two exits, where a counter was located at km 109. 
For the ground truth we used an AADTT that we computed from all available data for the section as the average of a count of 21 days in 2017 ($AADTT_{true, \ 2017}=2081$) 
and 182 days in 2015 ($AADTT_{true, \ 2015}=3427$) 
as reported by DNIT \citep{br_data}.
We also retrieved geospatial data of roads in Brazil from DNIT \citep{DNIT_shape}. These data are centered in one of the lanes, which is why we needed to expand the range used in the road filter to 40 m to ensure that both lanes pass the filter. This could result in errors if trucks were parked close to the road. 
We assumed a mean speed of 90 km/hr.

\subsubsection{Experimental procedure}
We fine-tuned the SSD model on images from Brazil (Appendix \ref{sec:detailDet}), and tested the performance on three different images of the same BR-116 section, using traffic variation factors trained on count data from the NY Thruway, California, and Germany.
We compared this to the performance when using the detection model and parameter settings from the NY Thruway.  

\subsubsection{Test results and discussion}

\begin{figure}[h]
\centering
   \includegraphics[scale=0.6]{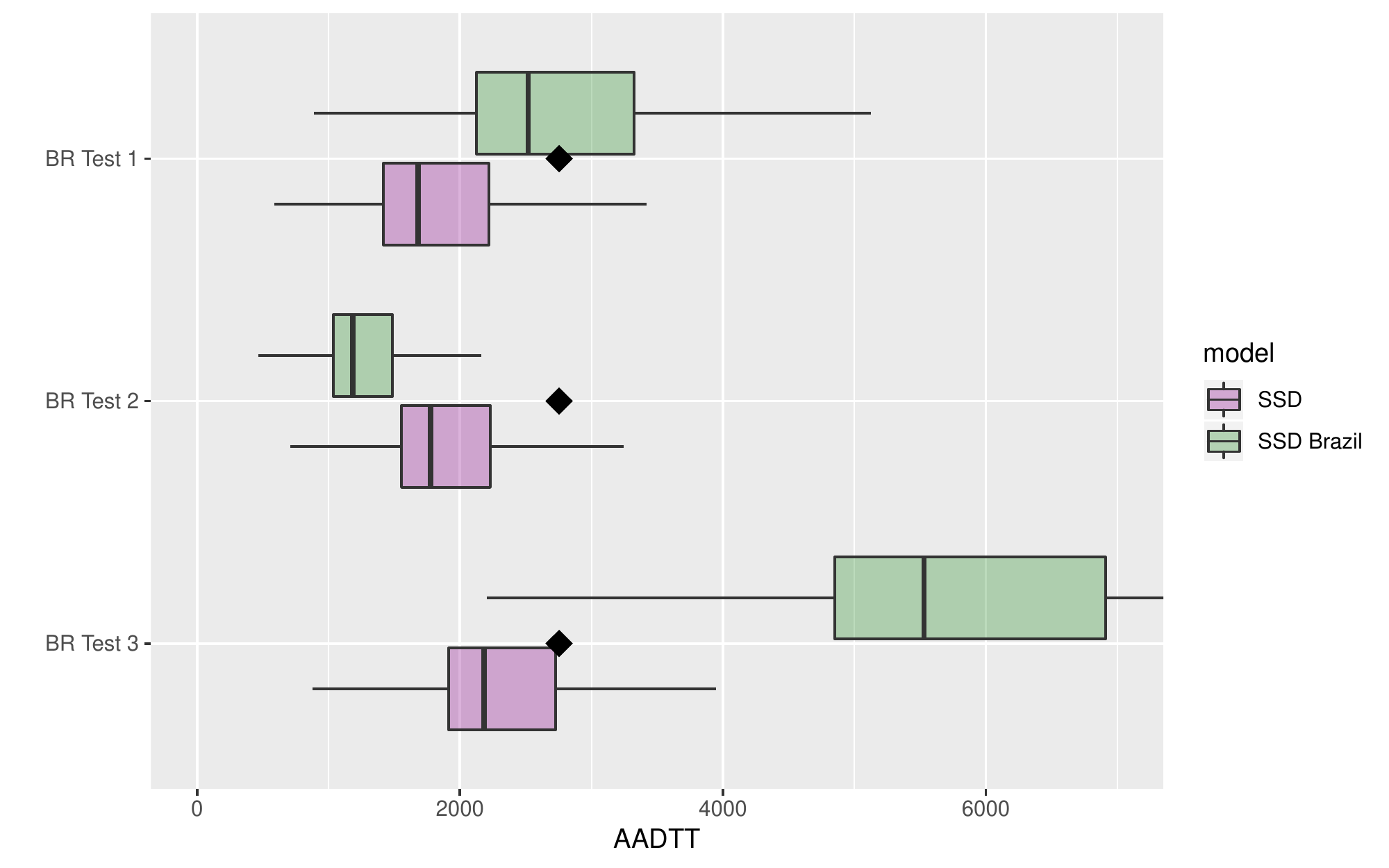} 
\caption{Predicted AADTT from satellite images (box plots) and ground-based AADTT (diamond) for a test section on BR-116 
at 3 different time stamps. We can see that the detection model trained on images from Northeastern US (SSD) underpredicts the AADTT but shows less variability than the model fine-tuned on images from Brazil (SSD Brazil). 
}
\label{fig:results}
\end{figure}

\begin{figure}[h]
\centering
   \includegraphics[width=0.9\linewidth]{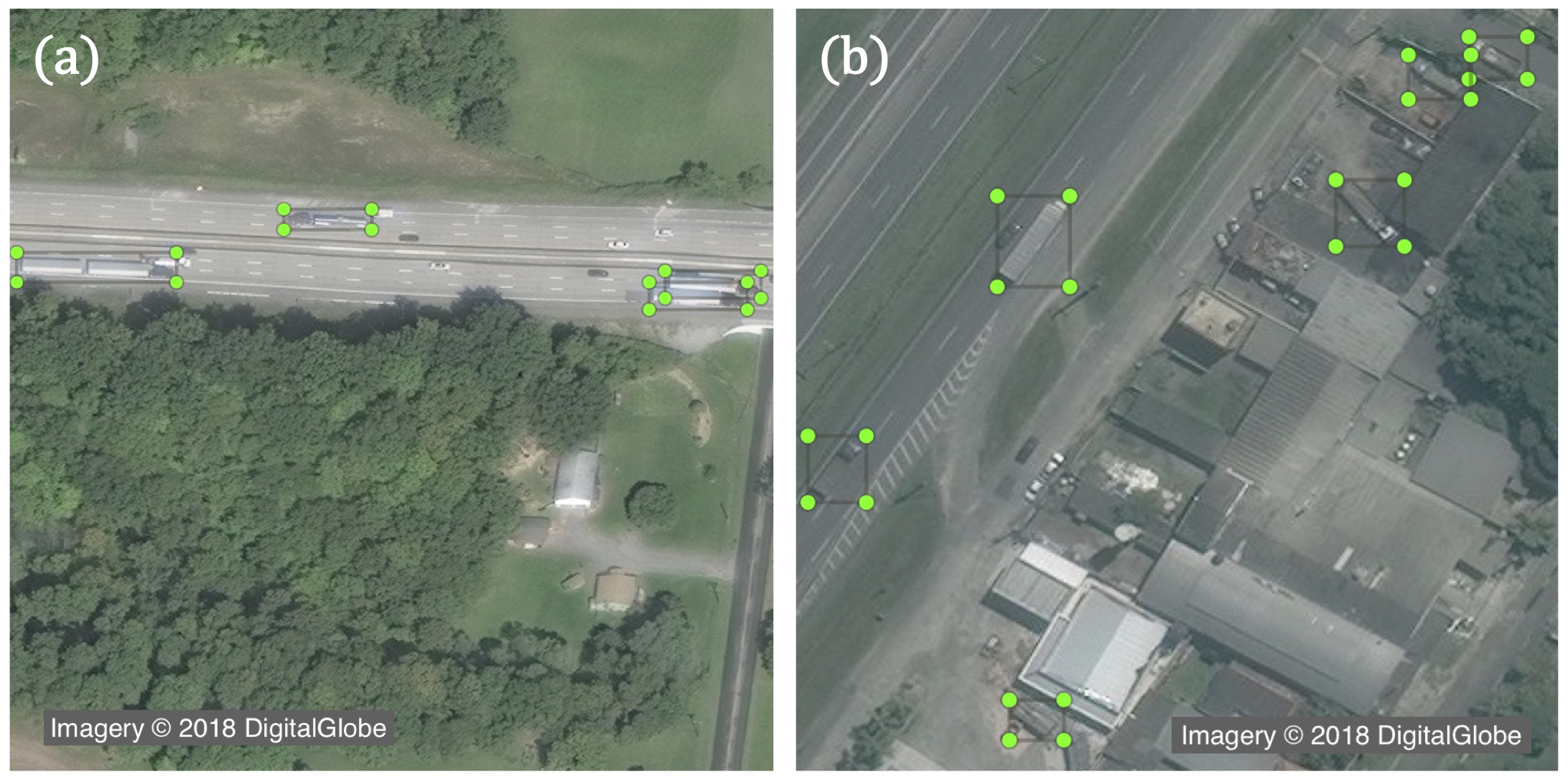}
\caption{Example images from NY Thruway (a) and BR-116 (b). Green boxes indicate annotated examples. 
While the geography looks similar in these images, 
the detection model did not generalize well between the countries (Fig.~\ref{fig:detecTest}). These images show that trucks seem to look different, which makes it hard for the model trained on US trucks to detect those in Brazil. The tractor is more box-shaped in Brazil.}
\label{fig:NYBR}
\end{figure}

We found that the US-trained detection model largely underpredicted the number of trucks in the new images, thus resulting in an underprediction of AADTT in all test cases (Fig.~\ref{fig:results}).  
The truck detection model trained on Brazilian images performed somewhat better (Fig.~\ref{fig:detecTest} and Appendix \ref{sec:appBR}), in particular for one of the test cases. 
These results indicate that additional fine-tuning of the detection model on images of the new location was necessary as new truck types occurred that were specific to Brazil and were not contained in the training dataset (Fig.~\ref{fig:NYBR}). 
However, we observed a much larger variability in the predicted AADTT when using the fine-tuned detection model (Fig.~\ref{fig:results}).

\section{Discussion and Conclusion}
\label{sec:discussion}

We find that we can use machine learning to count trucks in satellite images with reasonable accuracy. Using models of highway traffic patterns that were trained on data in other regions, a snapshot image can yield predictions of average annual daily traffic volumes that are acceptable, given the data limitations. Results could potentially be improved by using multiple satellite images taken of the same section at different times. 

While these initial results are promising, the uncertainties of the results are fairly large. Large estimation errors are not uncommon for predictions in energy and transportation, including for well-established conventional methods, and often the uncertainty of the estimate is not communicated. 
For example, when the U.S.~Bureau of Transportation Statistics updated its method to calculate total national road freight activity, the values increased by more than one fourth with the new method \citep{btstonmi}. Yet, as this is the only such estimate on road freight activity in the US, it is likely that policy makers routinely make decisions based on those point estimates. Our method is intended for regions where to date no AADTT data exist, and could provide critical decision support for governments in developing countries. The method currently still requires access to images, knowledge, and computing resources that might be difficult for some countries, but this could change in the near future. 

From our tests in the US and Brazil, we found that distinct truck types (rather than geography) can impact the prediction accuracy of the detection model, and additional training seems necessary to transfer the model between countries. Information on local driving patterns and labor laws could reduce the estimation error from the traffic monitoring model.

Future studies should focus on expanding validation and testing datasets. 
An internationally consistent database of traffic counts would also be helpful for building traffic monitoring models, as well as an extended study of how they transfer between countries.

\section*{Acknowledgements}
This work was supported by the Center for Climate and Energy Decision Making through a cooperative agreement between the National Science Foundation and Carnegie Mellon University (SES-0949710). We thank DigitalGlobe, Inc., for providing satellite image data. We are also grateful for computing resources that were made available through Google Cloud Platform and Microsoft AI for Earth, and for assistance from Francisco Ralston Fonseca with Brazilian data. We would like to thank Jay Apt, In\^es Azevedo, Parth \mbox{Vaishnav}, Patrick McSharry, and four anonymous reviewers for helpful comments on the draft of the paper.

\bibliographystyle{plainnat}
\bibliography{trucks_count_lib}

\newpage
\appendix

\section{Data preparation}

\subsection{Satellite images}

We use satellite images from DigitalGlobe, Inc., which are taken frequently and by a number of different satellites. We work with 3-channel RGB images of the satellite "World View 3" (VW03), as it has the highest resolution of 31 cm. We found that identifying vehicles in images of the other satellites was difficult. We attempted to select images with nearly no cloud-cover in relevant areas.

For training, we used images of several regions in the Northeastern United States. These include images of the NY Thruway but also other highways, and the regions around the highways such as parking lots and logistics centers. We fine-tune the model in another analysis step with images from highways in Brazil. For validation, we use images from the NY Thruway, and for testing we use images from the NY Thruway, California, and Brazil. Table \ref{tab:numberImages} shows the number of images that went into each analysis step. Each dataset contains a unique set of satellite images.

\begin{table}[h]
   \caption{Training, test, and validation data sizes for detection model. Non of the satellite images have been used in more than one dataset.}
   \centering
  \begin{tabular}{lllll} 
    \toprule
    Dataset & Satellite im. & Tiles & Number of trucks \\
    \midrule
   Training US & 14  & $300\times300$px & 2050   \\
   Training BR & 3  & $300\times300$px & 2027   \\
   Validation US & 4  & 12 larger tiles  & 340 (88 on road)   \\
   Validation BR & 2 & 10 larger tiles  & 409 (119 on road)   \\
    Test US & 14 & various tiles  & 541 on road  \\
    Test BR & 3 & various tiles  & 95 on road  \\
    \bottomrule
 \end{tabular}
 \label{tab:numberImages}
\end{table}

\subsection{Annotations}\label{sec:ann}
We used the Python-based annotation software "LabelImg" \citep{labelimg} to label the more than 5000 truck examples.
We marked each truck with a bounding box and a class label "Truck." Below we describe in detail, which types of vehicles we included as trucks. For cloudy images, we also annotated those trucks that are hardly visible through the cloud or in the shade of the cloud. The training data cover a wide variety of street orientations, and we used random horizontal flip during training, which ensured that the model learns a number of different vehicle orientations.

For the training data, we labeled large $3000\times3000$ pixel images, from which we created $300\times300$ pixel chips, where we only retained chips with truck examples. Note that this procedure reduces the number of truck examples somewhat with respect to the large images, as bounding boxes are cut and those examples are lost. 
We chose truck examples conservatively and prioritized accuracy of labeled examples over labeling as many as possible. This means, when in doubt, we chose not to label the vehicle, unless there are very obvious or interesting truck examples in the immediate vicinity. We have sometimes omitted parking areas and junk yards, where vehicles were parked close together. These examples were less useful for learning on highways but might have helped with images of dense traffic. We have excluded dense traffic from our proof of concept as it is less representative of intercity truck traffic. 

In contrast, for test images it was important to carefully label all likely trucks including those that were partially obstructed through trees, bridges etc.

\begin{figure}[h!]
\centering
   \includegraphics[width=0.80\linewidth]{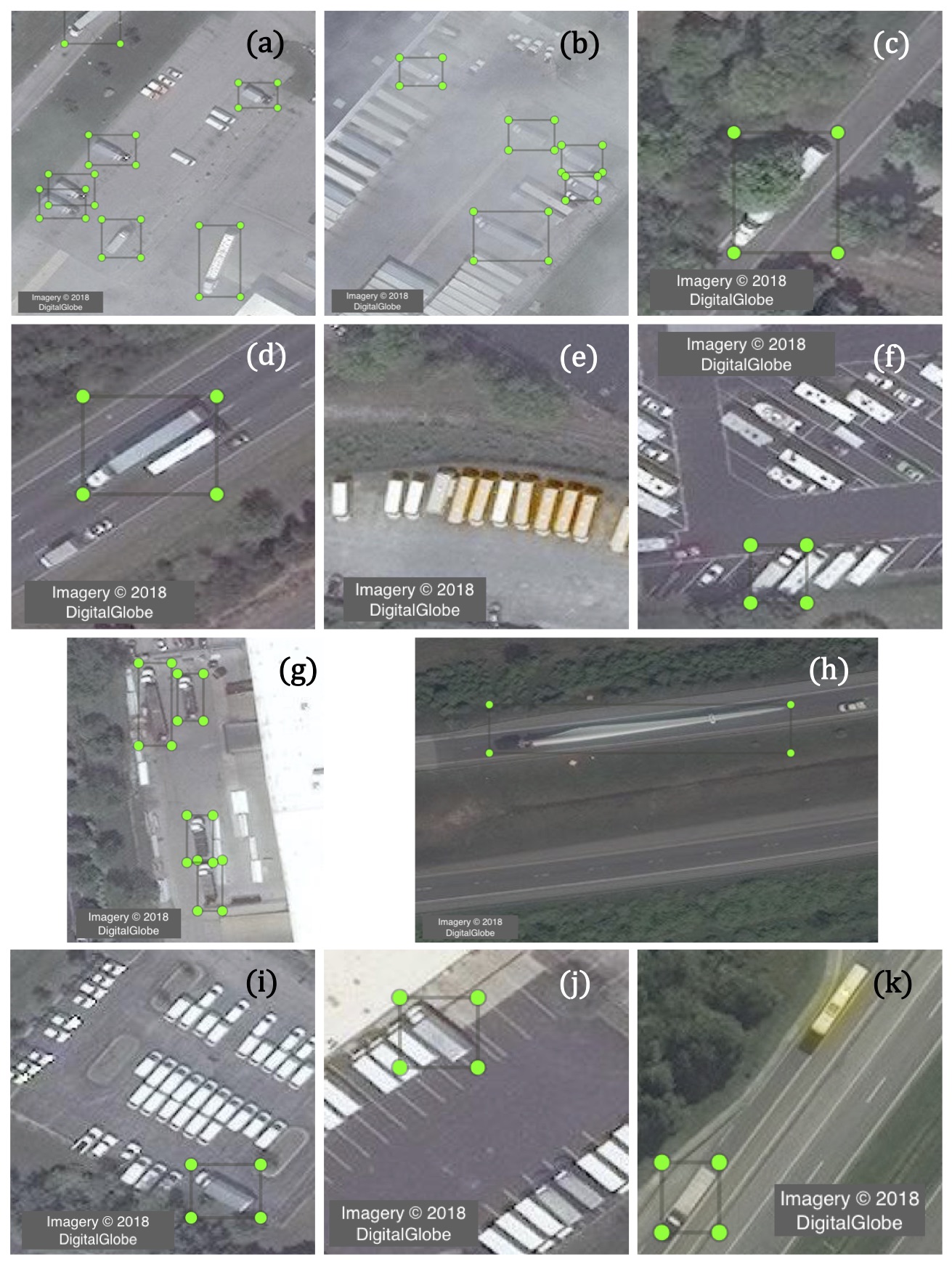} 
\caption{Image chips that illustrate what was labeled as a "Truck" indicated by a bounding box. Imagery \textcopyright~2018 DigitalGlobe, Inc. (a) Various trucks, (b) cloud cover reducing visibility, (c) truck obstructed by tree, (d) truck, RV and bus, (e) school buses, (f) RVs, (g) flatbeds, (h) example of special load: wind turbine blade, (i) vans, (j) smaller trailers, (k) yellow bus in Brazil.}
\label{fig:truckEx}
\end{figure}

We only annotated medium-duty trucks, and semi trucks with a trailer, including car carrier trailers, flatbed trucks or oversized transports such as wind turbine blades. We did not annotate pickup trucks, even if they pulled a trailer, and omitted vans, buses, caravans, and RVs. We also did not label tractors or trailers separately, only the combination, but also if the trailer was empty. 
Fig.~\ref{fig:truckEx} (a) 
shows an example of easily identifiable trucks of different sizes. The examples in Fig.~\ref{fig:truckEx} (b) and (c) 
are partially obstructed by clouds or trees but they can still be identified as trucks. Buses and RVs can easily be confused with trucks. The example in Fig.~\ref{fig:truckEx} (d) 
shows a bus (or a long RV) and something that is likely an RV that is pulling a car, both of which could be confused for a truck. Fig.~\ref{fig:truckEx} (e) and (f) 
show a number of yellow school buses and more RVs. We also included trucks with special trailers such as flatbeds (Fig.~\ref{fig:truckEx} (g)),  
or oversized load (Fig.~\ref{fig:truckEx} (h)). 
Smaller trucks and their similarity with vans were particularly difficult (Fig.~\ref{fig:truckEx} (i)).  
We considered everything that had a box that was elevated from the driver's cabin as a truck. Also, there were many examples of small parked trailers that had a white attachment, which could have also been a small driver's cabin (Fig.~\ref{fig:truckEx} (j)).  
For trucks that were docked to a building, we excluded those where only the trailer was visible but included those, where the tractor was still attached (Fig.~\ref{fig:truckEx} (b)). 
 
In the images from Brazil, we found what seemed to be yellow and white buses (Fig.~\ref{fig:truckEx} (k)), which appeared in multiple locations on the highway.  
After confirming with Google Street View that such buses frequently travel that highway, we have not labeled these as trucks.

\subsection{Vehicle counts}\label{sec:mon}

We used count data for four different regions. Those datasets comprised all counting stations with vehicle class distinction in California, Brazil, and Germany, and toll data from the NY Thruway. We only obtained counts for highways (or freeways), not smaller roads. Here, we describe each dataset and the respective data preparation, and conclude with a summary. To balance the training data between the regions, we used a sampling method that is explained as well.

\subsubsection{NY Thruway}

\begin{figure}
\centering
   \includegraphics[width=0.65\linewidth]{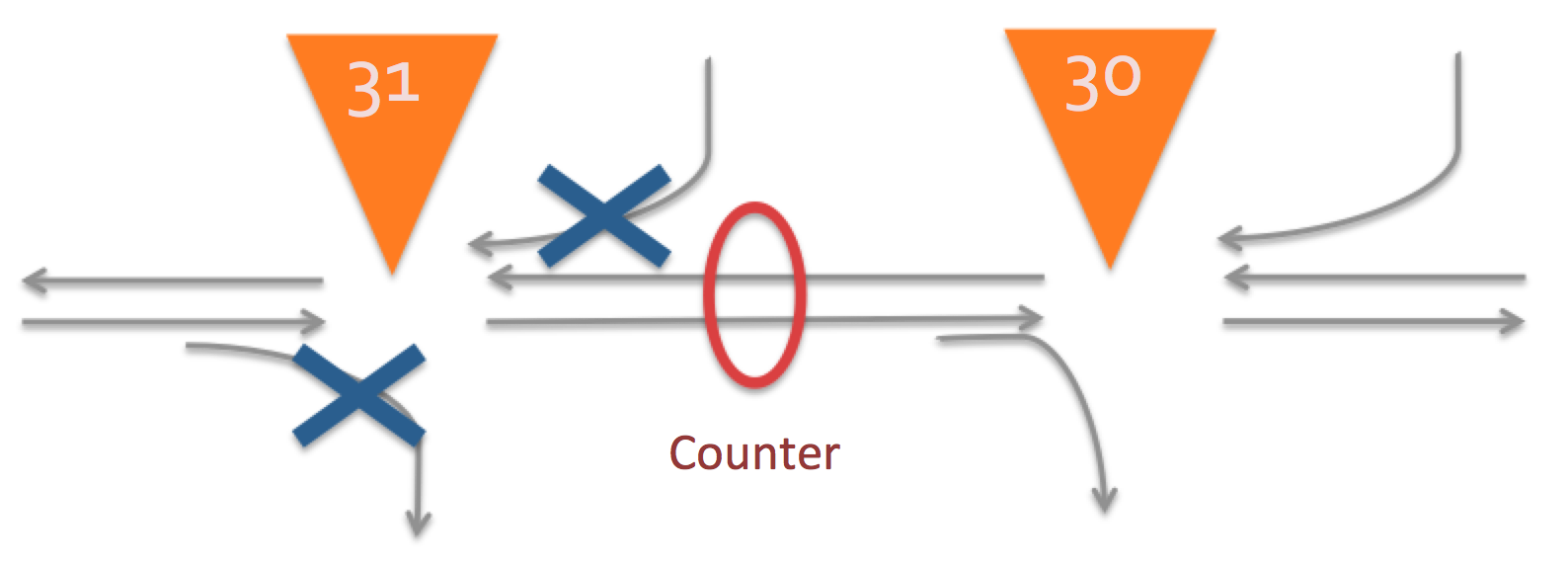} 
\caption{Schematic illustration of how we computed the NY Thruway counts from toll data. The orange cones indicate highway exits, and the grey lines are road sections. We did not consider those vehicle trajectories that are crossed out.}
\label{fig:countGT}
\end{figure}

We use toll data for the NY Thruway from 2016 for training and from 2017 for testing the whole model \citep{ezpass2017hourly, ezpass2016hourly}. The datasets contain the entrance and exit locations for every vehicle and the time it has entered the Thruway as recorded in the toll collection system. We considered all high vehicles with 3 or more axles to be trucks. To determine when a vehicle has passed a location between entrance and exit, we needed the speed it has traveled and the distance between highway exits. We assumed that every vehicle traveled 65 mi/hr, and we used Thruway mileposts available through the State of New York Thruway Authority \citep{thru_mileposts}. We determined the hourly counts by summing up all the vehicles that have entered a section between two highway exits within one hour. For example, for the stretch of road between Exit 30 and 31 we determined the hourly counts by summing up the number of vehicles that pass Exit 30 in one direction and pass Exit 31 in the other direction within that hour (see also Fig.~\ref{fig:countGT}). 

\subsubsection{California}

We obtained hourly count data through the California Department of Transportation (Caltrans) Performance Measurement System (PeMS), where we used hourly truck counts from the Caltrans Traffic Census Program for 2016 \citep{pemsTruck}. We furthermore used location information of census stations (weight in motion stations) from PeMS and Caltrans.
We also obtained AADTT information from Caltrans for a number of road segments in California \citep{caltransAADTT}.

\paragraph{Hourly data preparation.}
The CA datasets contain many short-term and some nearly continuous counters. We only used those counters that provide information on both directions, and added those directions up as well as the counts for all lanes. We dropped rows with vehicle classes $(0, 2, 3, 4, 15)$, as those include vehicles that are too small, or indicate malfunctions of the system.

\paragraph{AADTT check.} To ensure that we computed the AADTT correctly from the hourly data, we compared some estimates to those aggregate values provided through \citep{caltransAADTT}. Comparing counting locations in California proved difficult, as precise geographical information was not given. For example, we compared the value for census station 62010, which is weight in motion (WIM) location 73 for Caltrans or {\it rte. 5, district 6, leg A, JCT.RTE.43} in the 2016 table in \citep{caltransAADTT}. Here, we computed an AADTT of 8578 with the simple AADTT method, and Caltrans gives 8819. We ensured approximate compatibility also for other count stations. For the tests, we used the AADTT values from \citep{caltransAADTT}.

\subsubsection{Brazil}
The data for Brazil were made available through the Departamento Nacional de Infraestrutura de Transportes (DNIT) \citep{br_data}. The Brazilian dataset contains short-term as well as continuous counters on Brazilian national highways for several years. We worked with counts for the year 2017 for training the monitoring model because this was the year that had most data available.

\subsubsection{Germany}
The German agency Bundesanstalt f\"ur Stra\ss enwesen publishes hourly count data for continuous count stations on highways (Autobahnen) \citep{bast2018Germany}. We removed those stations that were faulty, indicated by an AADTT of 0. For training the monitoring model, we used the most recent data from 2017.

\begin{figure}
\centering
   \includegraphics[width=\linewidth]{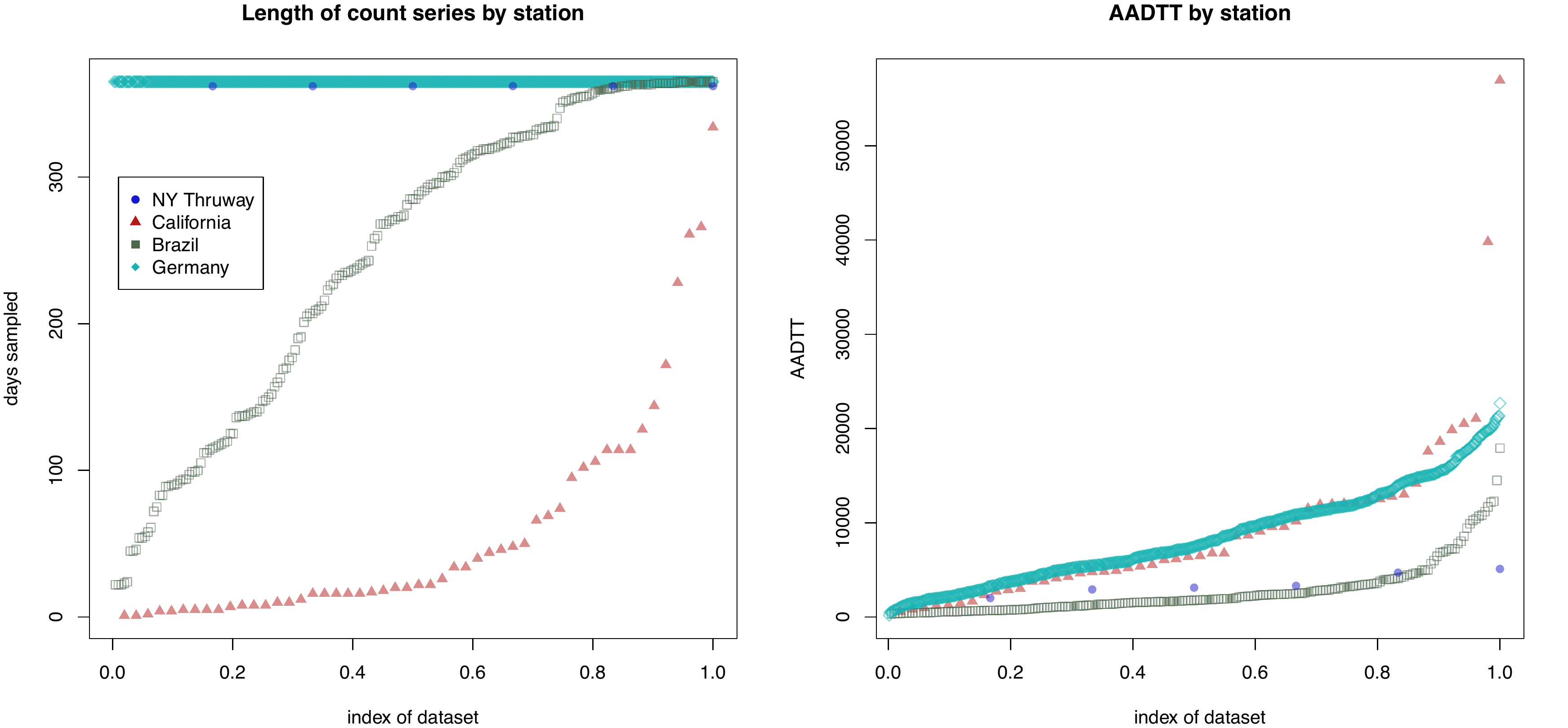} 
\caption{The count data vary between the regions. For example, German data contain only continuous counters, while for California mainly short-term counters are included. Filled points indicate those that where selected for the training dataset by the sampling procedure. We see that for the NY Thruway and California all count stations were used for training.}	
\label{fig:countsumm}
\end{figure}

\subsubsection{Count data summary and sampling}\label{sec:countsumm}
To balance between the regions, and ensure sufficiently high AADTT in the training data, we developed our own sampling procedure. We sampled from the stations with the longest count series (most "complete" stations) first, and then sampled from a selection of those with the highest AADTT, to arrive at a dataset of 10 continuous counting stations. As some of the datasets contain short-term counters, we iteratively increased the number of stations sampled until we arrived at a dataset with a number of points equivalent to 10 continuous counting stations.

The two plots in Fig.~\ref{fig:countsumm} show how complete the count time series are by station and the size of the AADTT, respectively. We see that the number of count stations as well as the length of count series by station differ considerably between the four regions. The figures also show those stations that were randomly selected to be part of the training dataset for the monitoring model. Count stations for test images were excluded.

\section{Experiment details}

\subsection{Computing the AADTT}\label{sec:detailMon}
There are several methods to compute the AADTT \citep{fhwa2016mon}. A simple method computes the average hourly count over all hours in the dataset (ideally a year) and multiplies by 24 to obtain the average daily value. A second method, the AASHTO method, computes average values for every weekday in a month, and then averages over these daily values. This method improves inter-year comparison, as it ensures that every annual value is computed with equal weights between the weekdays. 
As we had a lot of incomplete and short-term counts in our datasets, we used the first (simple) method to compute the AADTT.

\subsection{NY Thruway truck detection model}\label{sec:detailDet}
\begin{figure}[h]
\centering
   \includegraphics[scale=0.6]{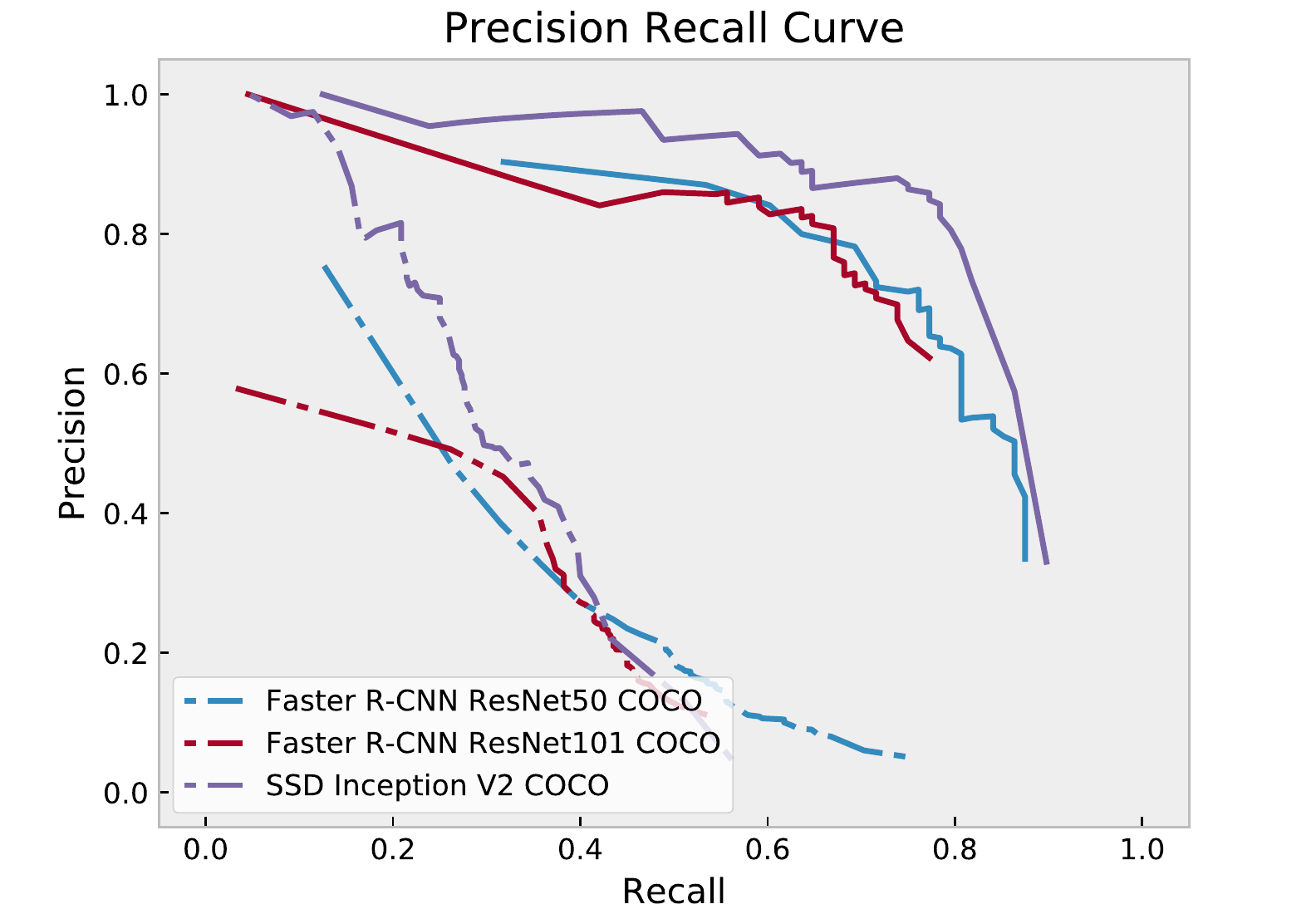}
\caption{Precision-recall curves for validation images on full image (dashed) and constrained to the road (solid). All of the models performed better when used for on-road predictions, as those often contain less difficult examples.}
\label{fig:pretrain}
\end{figure}

\paragraph{Validation details.}
We validated the models using precision and recall 
on the validation dataset for the whole image, and for the subset of trucks that are on the road. The model with the lowest count error can achieve simultaneously the largest precision and recall. 
We computed the average precision and recall over all validation images together, and did not average the performance over each image separately. The optimal values are reported in Table~\ref{tab:results}.
From Table~\ref{tab:results} and Fig.~\ref{fig:pretrain}, we can see that the models performed better when the experiment was constrained to the road. The full image can contain more difficult examples, for example clustered trucks on parking lots or less typical trucks in junk yards or construction sites. 

\begin{table}[h]
   \caption{Performance on road for optimal prediction probability $p_{pred}$; pre-trained on COCO and fine-tuned on $\sim 2000$ trucks.}
   \centering
  \begin{tabular}{llllll} 
    \toprule
    Faster R-CNN &  Min. Count Error & $p_{pred}$ & Precision   & Recall \\
    \midrule
   ResNet50 & 0.205 & 0.800 & 0.773 & 0.694  \\
    ResNet101 &  0.216 & 0.320  & 0.693  & 0.735  \\
    SSD Inception V2& 0.170 & 0.155  & 0.773  & 0.859  \\
    \bottomrule
 \end{tabular}
 \label{tab:results}
\end{table}

\paragraph{Bias in count error.}

\begin{figure}
\centering
   \includegraphics[scale=0.6]{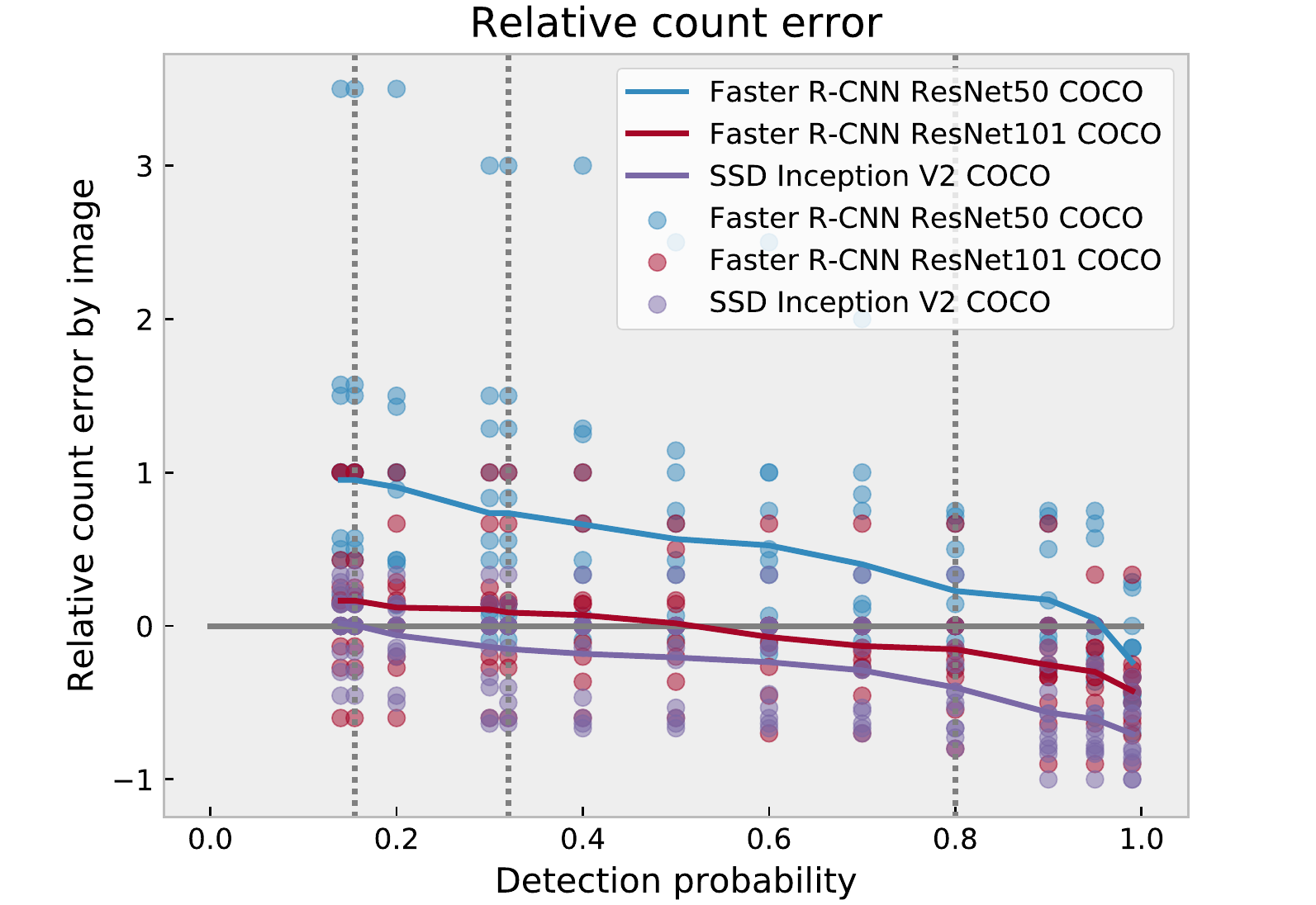} 
\caption{Count errors as a fraction of the true number of annotated trucks in each image on the road (points). Negative values indicate a lower predicted number of trucks than the number of annotated trucks. The lines indicate the mean over all images per model. We see that at the optimal prediction probability (grey dashed lines), the SSD is not biased.}	
\label{fig:scatter}
\end{figure}

We were interested in understanding if the count error score we use was biased. Fig.~\ref{fig:scatter} shows the individual relative count errors for each validation image with a negative value corresponding to underprediction. The mean over all images is around zero for the optimal prediction probability of the SSD, but both Faster R-CNN detectors are overpredicting for their respective optima. The SSD, however, tends to underpredict for the widest range of probability thresholds. This indicates that it might be useful to further explore if lower count errors could be achieved with other object detection models.

\subsection{Brazil truck detection model}\label{sec:appBR}

\begin{figure}
\centering
   \includegraphics[scale=0.6]{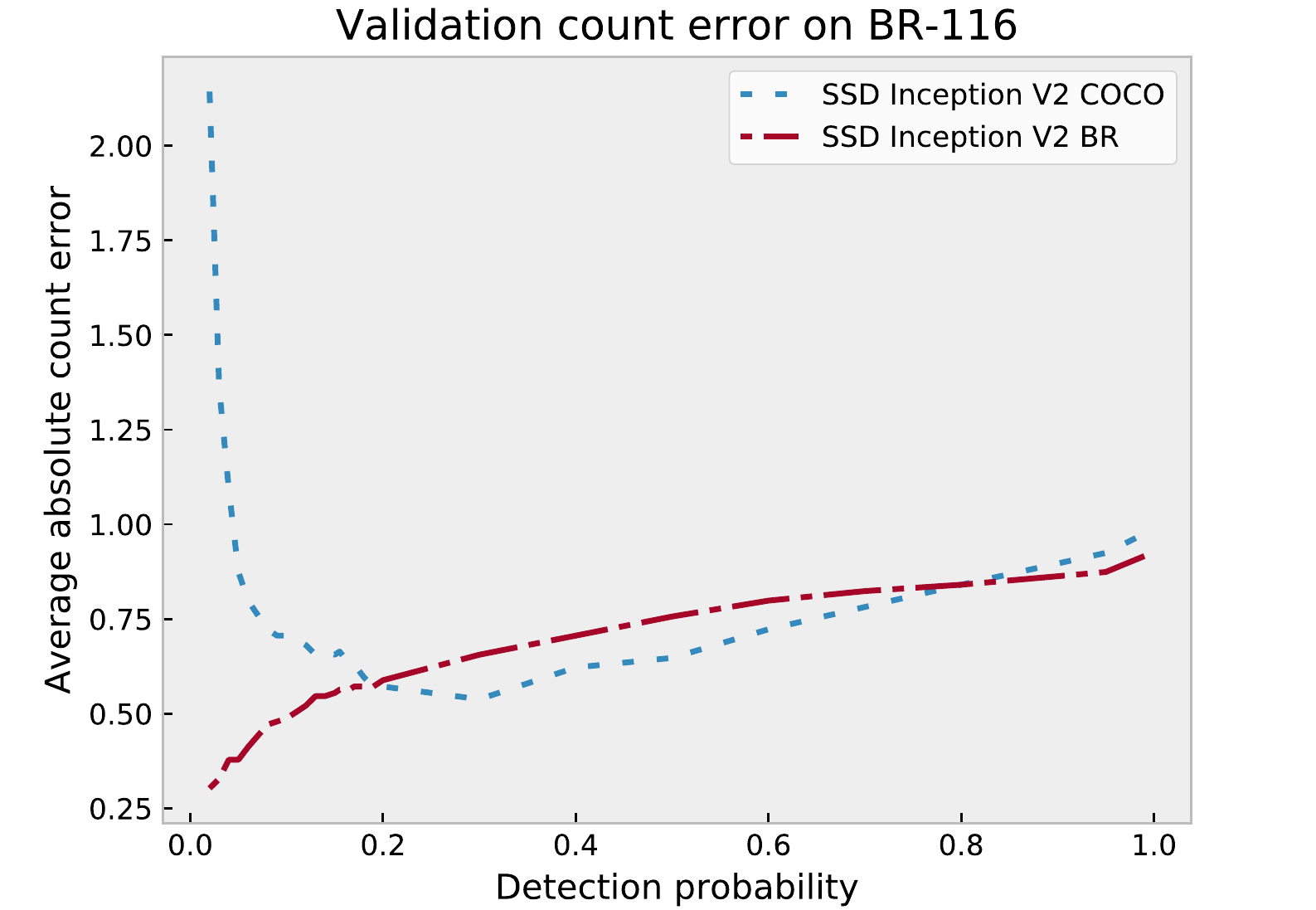} 
\caption{Count errors for the validation set for Brazil. The second model is fine tuned on training images from Brazil and outperforms the model trained on images from Northeastern US.}
\label{fig:countBR}
\end{figure}

The validation count curves are shown in Fig.~\ref{fig:countBR}. We see that for very low probability thresholds the model that is fine-tuned on local data outperforms the detection model that has only seen images from the US. 
For the test cases we compare the original model with original parameter settings, and the fine-tuned and validated detection model with $p=0.02$.

\end{document}